\documentclass[%
aps,prl, twocolumn,superscriptaddress, amsmath,amssymb,
]{revtex4-1}

\usepackage{placeins}
\usepackage{graphicx}
\usepackage{bm}
\usepackage{xcolor}

\begin{document}

\title{Laser Cooling of Optically Trapped Molecules}

\author{Lo\"ic Anderegg}
\email{anderegg@g.harvard.edu} 
\author{Benjamin L. Augenbraun}
\author{Yicheng Bao}
\author{Sean Burchesky}
\author{Lawrence W. Cheuk}
\affiliation{Department of Physics, Harvard University, Cambridge, MA 02138, USA}
\affiliation{Harvard-MIT Center for Ultracold Atoms, Cambridge, MA 02138, USA}

\author{Wolfgang Ketterle}
\affiliation{Harvard-MIT Center for Ultracold Atoms, Cambridge, MA 02138, USA}
\affiliation{Department of Physics, Massachusetts Institute of Technology, Cambridge, MA 02139, USA }

\author{John M. Doyle} 
\affiliation{Department of Physics, Harvard University, Cambridge, MA 02138, USA}
\affiliation{Harvard-MIT Center for Ultracold Atoms, Cambridge, MA 02138, USA}

\date{\today}

\begin{abstract}

Calcium monofluoride (CaF) molecules are loaded into an optical dipole trap (ODT) and subsequently laser cooled within the trap. Starting with magneto-optical trapping, we sub-Doppler cool CaF and then load $150(30)$ CaF molecules into an ODT. Enhanced loading by a factor of five is obtained when sub-Doppler cooling light and trapping light are on simultaneously. For trapped molecules, we directly observe efficient sub-Doppler cooling to a temperature of $60(5)$\,$\mu\text{K}$. The trapped molecular density of $8(2)\times10^7$\,cm$^{-3}$ is an order of magnitude greater than in the initial sub-Doppler cooled sample. The trap lifetime of 750(40)\,ms is dominated by background gas collisions.

\end{abstract}

\maketitle

Wide-ranging potential applications have created interest in producing ultracold molecules. Heteronuclear bialkali molecules, assembled from ultracold atoms, have been brought near quantum degeneracy and have enabled the study of long-range dipolar interactions and quantum-state-controlled chemistry~\cite{Yan2013,Ospelkaus2010}. Other kinds of molecules are desirable for important applications~\cite{DiRosa2004,prehn2016optoelectrical, kozyryev2016lasercoolpoly}, such as quantum simulation~\cite{zoller06,carr09}, quantum information~\cite{demille02qi,yelin06}, and precision tests of fundamental physics~\cite{ACME14,hinds12,Lim2017,Kozyryev2017,carr09}. Polar molecules featuring unpaired electron spins (e.g. those with $^2\Sigma$ ground states) are sought after because they can possess both a non-zero electric and magnetic moment. They can be used to simulate a large variety of lattice spin models~\cite{pupillo08}, some of which host topological phases, opening up the possibility of topologically protected quantum memories and gates~\cite{zoller07}. A generic class of $X\,^2\Sigma$ molecules allows for optical cycling~\cite{DiRosa2004}, a crucial property that permits high fidelity optical detection and laser cooling to ultracold temperatures. Recently, laser cooling and magneto-optical trapping of CaF and SrF~\cite{barry14,norrgard16RF,steinecker16,Truppe2017subdoppler,anderegg17} were reported. Trapping in a conservative trap is needed for a wide variety of future work ranging from precision spectroscopy, degenerate quantum gases and quantum information, to evaporative cooling.

In this Letter, we demonstrate loading of laser-cooled CaF molecules into an optical dipole trap (ODT) and the sub-Doppler laser cooling of the optically trapped molecules. We start with a buffer gas cooled beam of CaF, followed by laser slowing, rf magneto-optical trapping, and then sub-Doppler cooling to 40\,$\mu$K. The presence of the sub-Doppler cooling light during the loading of the ODT significantly enhances the trapped number over loading in the absence of such light. 150(30) molecules are trapped in a 380(60)\,$\mu$K deep ODT operating at 1064\,nm with a density of $8(2)\times10^7$\,cm$^{-3}$. We directly demonstrate that laser cooling occurs in the presence of the trap light. Our approach is applicable to a large class of molecules including polyatomic species~\cite{kozyryev2016lasercoolpoly,kozyryev2017srohsisyphus} that cannot easily be assembled from ultracold atoms.

Conservative traps in ultracold atom experiments are commonly formed from magnetic or optical fields. Recently, magnetic trapping of laser cooled polar molecules was reported~\cite{Williams2017,McCarron2017}. The significant advantages of magnetic traps are large volumes and high trap depths, which open up the possibility of further evaporative or sympathetic cooling. A disadvantage is that only low-field seeking states may be trapped, which precludes trapping of the absolute ground state. These states are susceptible to spin-changing inelastic loss during collisional cooling. Magnetic traps also introduce large state-dependent energy shifts, which can render further laser cooling of trapped samples difficult, as well as prevent precision spectroscopy.

Optical traps, despite their much smaller trap volumes and lower trap depths, can offer nearly state-independent trapping potentials. This allows for laser cooling of trapped molecules, which, as demonstrated in this work, can lead to density enhancement. This opens up alternative paths to high phase space densities, which can occur on much shorter time scales and be more efficient than evaporative cooling. For example, laser cooling of optically trapped atoms has recently led to production of Bose-Einstein condensates without evaporation~\cite{Stellmer2013, Hu2017}. Another advantage of optical traps is that small-scale features on the order of the wavelength of the trapping light can be created. 
Although optical trapping of diatomic molecules have previously been reported~\cite{Takekoshi1998,Neyenhuis2012}, these molecules could not be laser cooled and were assembled from ultracold atoms.


The starting point of our experiment is a radio-frequency (rf) MOT of CaF molecules loaded from a cryogenic buffer gas beam. This setup has been described previously~\cite{anderegg17}, except that here we implement chirped slowing of the CaF molecular beam~\cite{Truppe2017}. The rf MOT is the same, operating on the X$^2 \Sigma^+(N=1) \rightarrow$ A$^2\Pi_{1/2}(J^\prime=1/2)$ transition, and consists of three retro-reflected MOT beams, along with lasers to repump out of the $v=1,2,3$ vibrational levels. In addition to forming the rf MOT, the MOT beams are also used for sub-Doppler cooling. Each MOT beam has a $1/e^2$ diameter of 9\,mm and contains 30\,mW of X($v=0$)$\rightarrow$A($v=0$) light and 30\,mW of X($v=1$)$\rightarrow$A($v=0$) light. The power distribution among hyperfine components, addressed with AOMs, for both X($v=0,1$)$\rightarrow$A($v=0$) are 18\%, 36\%, 36\%, and 10\% for the states $\lvert J, F \rangle = \lvert 3/2, 1 \rangle$,  $\lvert 3/2, 2 \rangle$, $\lvert 1/2, 1 \rangle$, and $\lvert 1/2, 0 \rangle$, respectively. When loading molecules into the rf MOT, the MOT beams are detuned from resonance by $\Delta_\text{MOT}= -2\pi \times 9$\,MHz (Fig.~1(b)). The MOT beams are initially held at full intensity for 15\,ms, capturing $10^5$ molecules into the MOT. The intensity of each MOT beam is then reduced by a factor of 8 over 30\,ms. Lower intensities reduce the sub-Doppler heating associated with red detuning on a $J\rightarrow J$ or $J\rightarrow J-1$ transition~\cite{Devlin2016} and decrease the MOT temperature from 2\,mK to 0.35\,mK while increasing the density to $5\times10^6$\,cm$^{-3}$.

\begin{figure}
\includegraphics[width=\columnwidth]{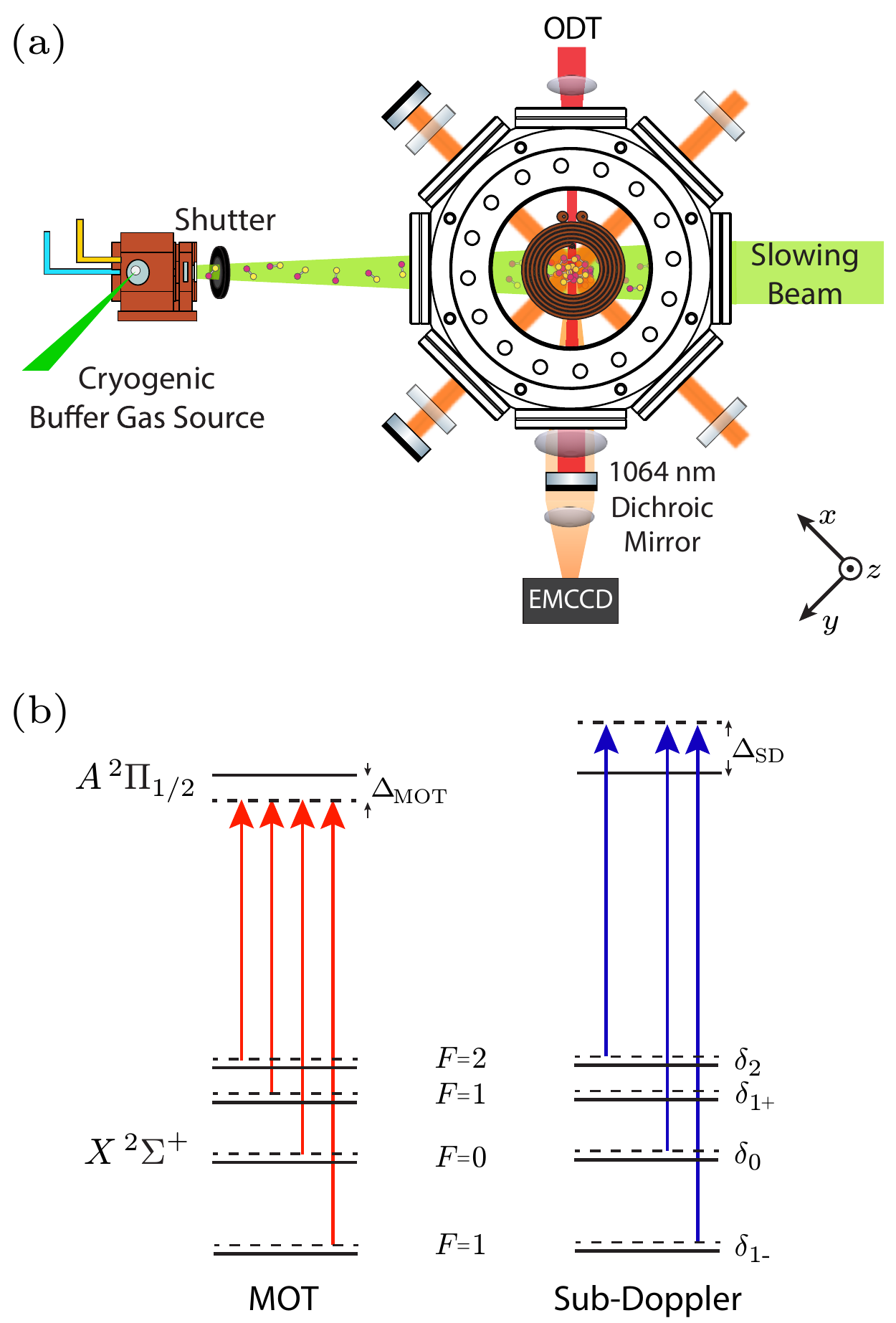}

        \caption{(a) Layout of the experimental apparatus. See \cite{anderegg17} for details. (b) Laser detunings for the rf MOT and for sub-Doppler cooling. For the rf MOT, the light is detuned by $\Delta_\text{MOT}=-2\pi \times 9$\,MHz; for sub-Doppler cooling, the light is detuned by $\Delta_\text{SD}=2\pi \times 28$\,MHz. The individual hyperfine sidebands are addressed with AOMs, and have detunings of $\{\delta_{2}, \delta_{1^+}, \delta_{0}, \delta_{1^-} \} = 2\pi \times \{1.8,0,0,-4.3\}$\,MHz. Positive detunings correspond to blue detunings relative to the respective resonances for the various hyperfine components.}
\end{figure}

The inverted angular momentum structure of CaF is similar to the $D1$-line in alkali atoms, and allows for further cooling at blue detunings~\cite{grynberg94,Sievers2015,Devlin2016, Truppe2017subdoppler}. To perform sub-Doppler cooling, the MOT beams and the MOT magnetic gradient are switched off in 200\,$\mu$s, during which time the laser is detuned to the blue, $\Delta_\text{SD}\approx +3\, \Gamma$. The MOT beams, with polarization switching turned off, are then switched back on at full intensity, but without the $J=3/2,F=1$ component (Fig.~1(b)), which becomes resonant with $\lvert 3/2, 2 \rangle$ when the detuning is $+3\,\Gamma$. Repumping out of $\lvert 3/2, 1 \rangle$ is still accomplished, albeit at a reduced rate by off-resonant light that nominally addresses the $\lvert 3/2, 2 \rangle$ state. For optimal cooling, compensation coils are used to cancel ambient magnetic fields to better than $0.1$\,G. In a time of $\sim\!\!\,100\,\mu$s, the molecules are cooled to 40\,$\mu$K, much lower than the Doppler cooling limit of $200\,\mu$K.

After sub-Doppler cooling, we load the molecules into a far-detuned optical dipole trap (ODT). The ODT is formed from 12.7\,W of single-frequency 1064\,nm light focused to a waist of $29\,\mu\text{m}$, which produces a trap with a calculated depth of 380(60)\,$\mu$K, with radial (axial) trap frequencies of $2\pi\times2.5$\,kHz\,($2\pi\times21$\,Hz). The ODT light is reflected by a dichroic mirror that transmits the fluorescence of the molecules, which is imaged onto an EMCCD camera. The reflected ODT beam is directed to a beam dump, away from the trapped molecules.

In order to capture molecules in the optical trap, the ODT light is switched on at the start of the sub-Doppler cooling. The sub-Doppler light is then ramped down in intensity by 30\% in the first 10\,ms and left on for a further 5\,ms. We wait for 50\,ms to allow untrapped molecules to fall from the imaging region, before imaging using a 0.5\,ms pulse of light resonant with the X$\rightarrow$A transition. With optimal parameters, 150(30) molecules are transferred into the ODT. For the trapped molecules, we measure a temperature of 60\,$\mu$K. Using the measured ODT beam profile and the number of trapped particles, we determine the peak density in the trap to be $8(2)\times10^7$~cm$^{-3}$, two orders of magnitude higher than recently reported densities of magnetically trapped $^2\Sigma$ molecules~\cite{Williams2017,McCarron2017}, and one order of magnitude higher than the initial magneto-optical trap. Our estimation is obtained by assuming thermally distributed molecules in the trapping potential of the ODT, and agrees with classical Monte Carlo simulations of particles drawn from a uniform density distribution at the experimentally measured sub-Doppler temperature. This corresponds to a peak phase space density of $2\times 10^{-9}$, three times higher than in the free-space sub-Doppler cooled cloud, and nearly three orders of magnitude higher than reported for magnetically trapped CaF molecules~\cite{Williams2017}.

\begin{figure}[t!]
\centering
\includegraphics[width=\columnwidth]{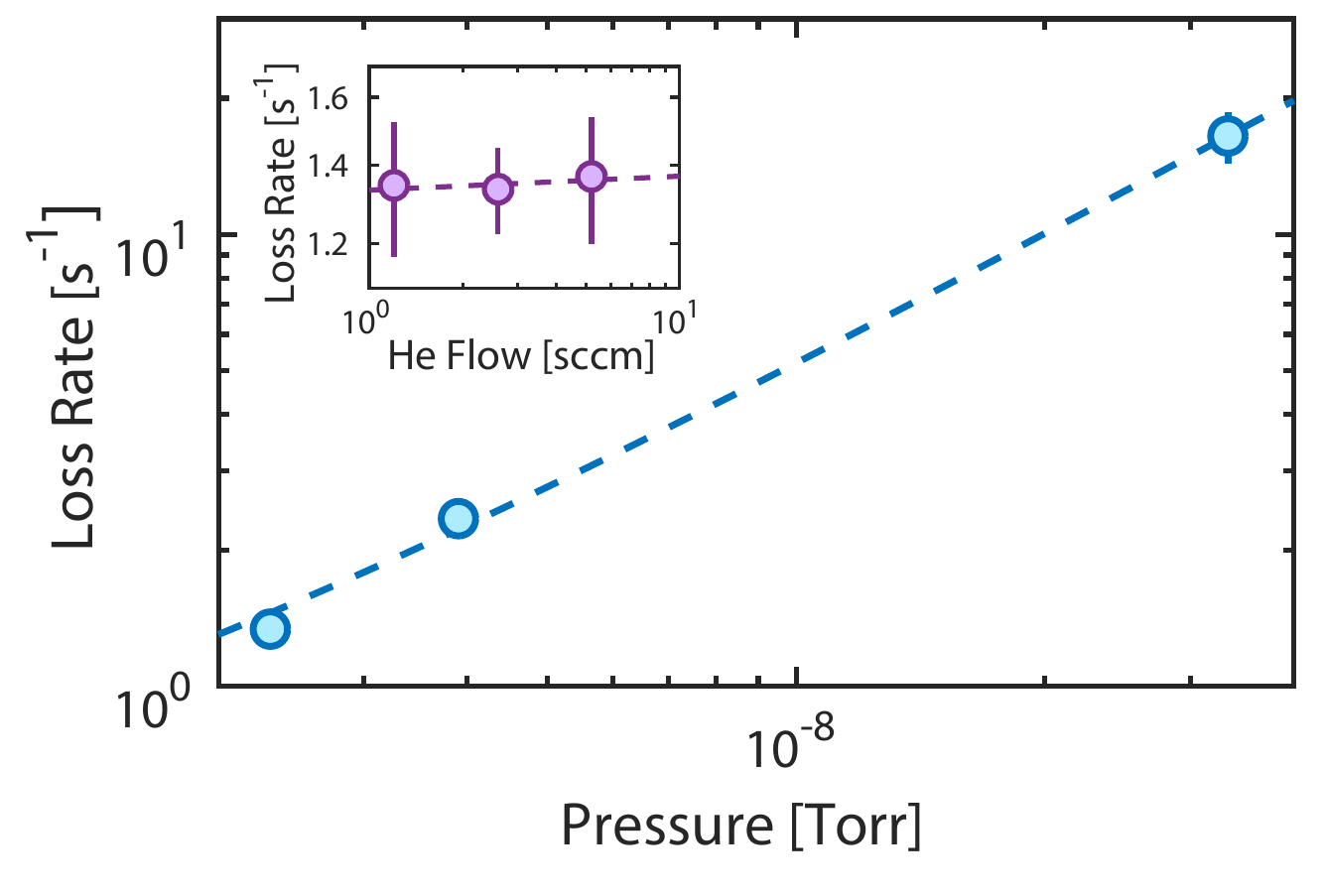} 
\caption{Loss rate of molecules trapped in the ODT. The inset shows the loss rate as a function of the buffer gas flow rate. We observe negligible effect of increasing the buffer gas flow, but significant dependence on the background pressure. The dependence of loss rate on buffer gas flow is measured to be 0.0(2)\,s$^{-1}$sccm$^{-1}$, while the dependence on background pressure is $5.5(1)\times 10^{8}$\,s$^{-1}$Torr$^{-1}$. Dashed lines show linear fits.
}
\label{fig:lifetime}
\end{figure}

An important characteristic of the ODT is the lifetime of trapped molecules, which is measured to be 750(40)\,ms ($1/e$ decay time), shorter than the calculated lifetime ($\gg$\,1\,s) due to heating from off-resonant photon scattering of the 1064\,nm ODT light. To determine what limits the lifetime, we vary the flow rate of helium into the buffer gas cell, with no appreciable effect on the lifetime (Fig.~\ref{fig:lifetime} inset). An in-vacuum UHV shutter that is open for $<\!\!\,10\,\text{ms}$ during each experimental cycle eliminates the effects of buffer gas collisions on the 1\,s timescale. To explore the dependence of loss rate on background pressure, we vary the MOT chamber pressure from $1\times10^{-9}$\,Torr to $3\times10^{-8}$\,Torr and find a dependence of $5.5(1) \times 10^8$\,s$^{-1}$Torr$^{-1}$ (Fig.~\ref{fig:lifetime}). This indicates that at our current operating conditions of $1\times10^{-9}$\,Torr, the loss rate of molecules from the ODT is dominated by collisions with background gas.

\begin{figure}[t!]
\centering
\includegraphics[width=\columnwidth]{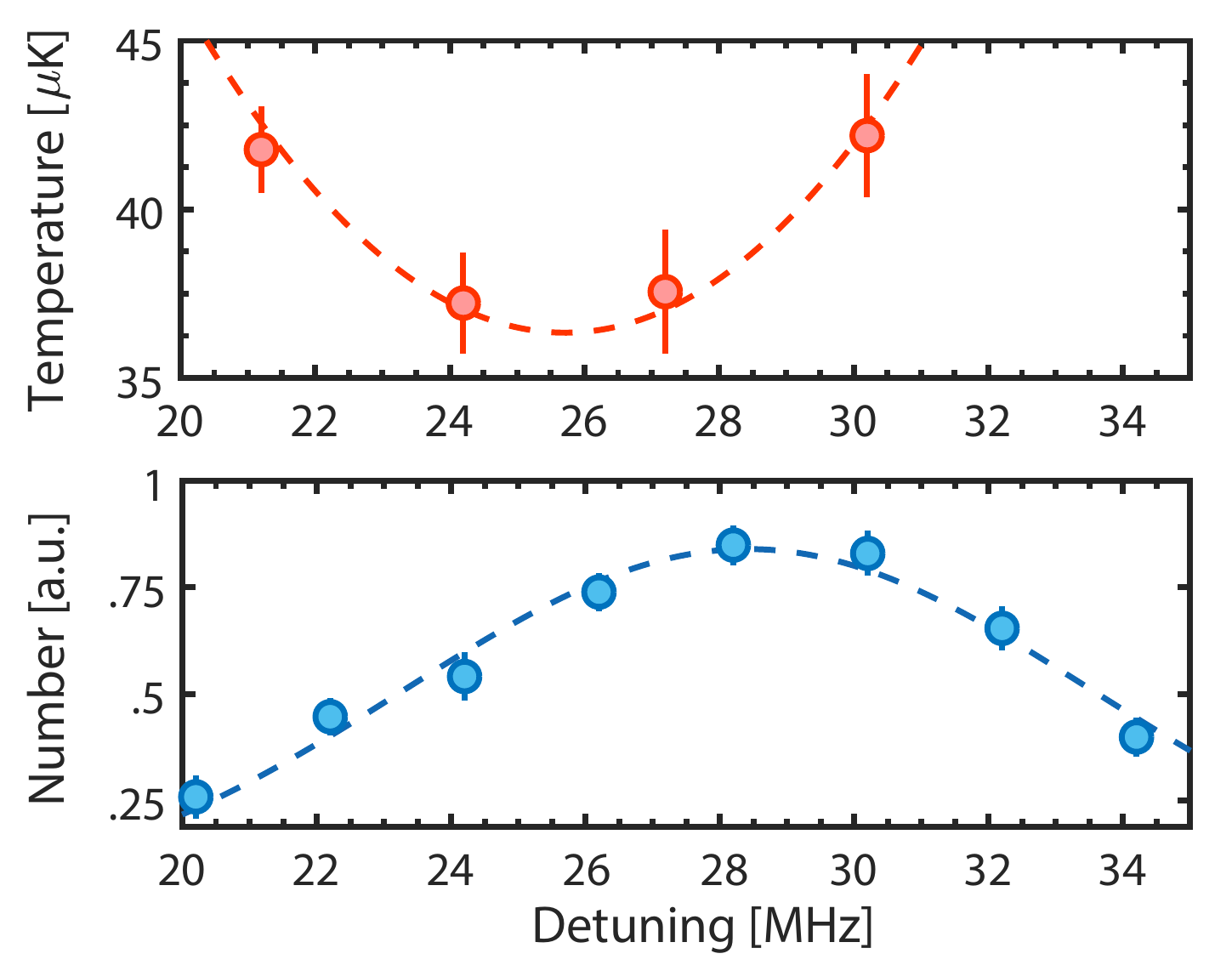} 
\caption{Dependence of sub-Doppler cooling and ODT loading on laser detuning. Top: Free-space sub-Doppler temperature as a function of detuning $\Delta_\text{SD}$. The dashed line shows a quadratic fit. Bottom: Number of trapped molecules in the ODT as a function of $\Delta_\text{SD}$. The dashed line shows a Gaussian fit. We observe a $3\,$MHz shift between optimal ODT loading and optimal free-space sub-Doppler cooling, which is consistent with the estimated ac Stark shift due to the ODT. }
\label{fig:stark}
\end{figure}

We further characterize the effect of the sub-Doppler light on the ODT loading process. We first vary the frequency of the sub-Doppler cooling light during ODT loading. As shown in Fig.~\ref{fig:stark}, optimal loading occurs when the sub-Doppler cooling light is detuned +3\,MHz relative to the free-space cooling frequency that produces the lowest temperature. The detuning is consistent with estimates of the ac Stark shift on the X$\rightarrow$A transition arising from the ODT.

\begin{figure}[t]
\centering
\includegraphics[width=\columnwidth]{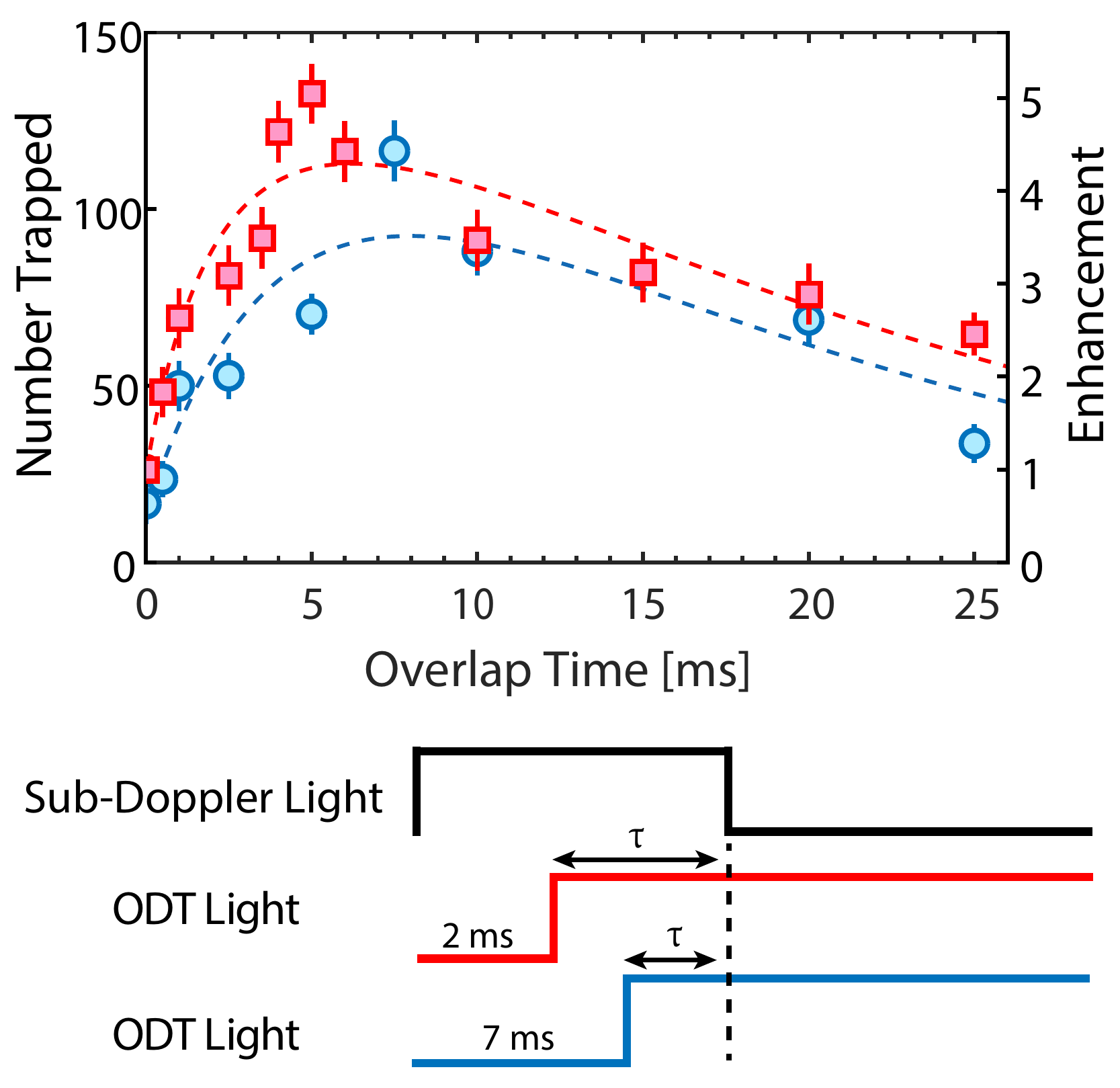} 
\caption{Loading of molecules into the ODT as a function of overlap time, $\tau$, with the sub-Doppler light. Shown in red squares (blue circles) is the number loaded when the cooling light is turned on 2\,ms (7\,ms) prior to the ODT light. The number of molecules loaded into the ODT is enhanced by up to a factor of 5 with sub-Doppler light. The enhancement is relative to $\tau=0$ for $2$\,ms of free-space cooling. The dashed lines indicate fits to a rate equation model, with constant loss rate and loading rate proportional to $\tau$. 
}
\label{fig:odttemp}
\end{figure}

We next vary the overlap duration of the cooling light and the ODT. This is accomplished as follows. We shorten the sub-Doppler intensity ramp to $2\,\text{ms}$ with the ODT off. Subsequently, the ODT is switched on, and we vary the amount of time that the sub-Doppler light overlaps with the ODT light. Zero temporal overlap corresponds to direct capture of sub-Doppler cooled molecules. As shown in Fig.~\ref{fig:odttemp}, we find that the number of trapped molecules increases by up to a factor of 5 and reaches a peak at $5\,\text{ms}$. This increase suggests that sub-Doppler cooling is effective even in the presence of the trap light. To verify that the enhancement is not due to additional sub-Doppler cooling in free space, we increase the duration of the free space sub-Doppler cooling by an additional $5\,$ms before switching on the ODT. With the additional free-space cooling, we find a smaller initial number of loaded molecules, likely due to a lower central density resulting from the longer sub-Doppler cooling time. 
Despite a smaller initial number, the number of loaded molecules again rises as a function of overlap time. The peak number is reached at $\sim\!\!\,10\,\text{ms}$, at a similar level to the case without extra sub-Doppler cooling. 

\begin{figure}[t!]
\centering
\includegraphics[width=\columnwidth]{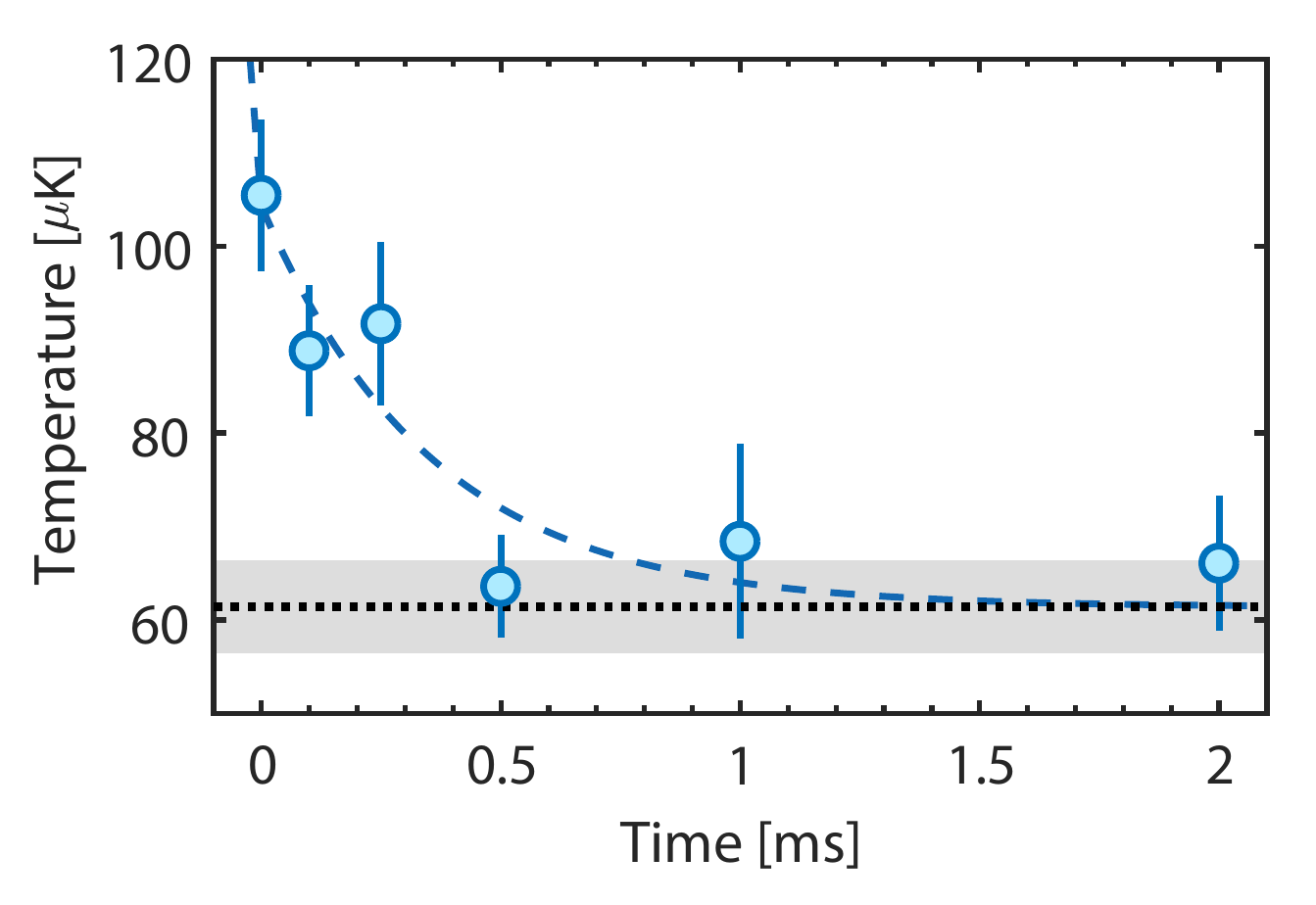} 
\caption{Cooling of optically trapped molecules. Trapped molecules are heated by a 40\,$\mu$s pulse of resonant light at time $t=0$. Shown is the temperature as a function of varying sub-Doppler cooling time beginning at $t=0.2\,\mu\text{s}$. An exponential fit, shown by the dashed blue line, yields a $1/e$ time constant of $\sim \!\!\,300\mu$s. The temperature prior to heating is indicated by the black dotted line.}
\label{fig:cooling}
\end{figure}

The initial rise at short times indicates that there is a time window of $\sim\!\!\,10\,\text{ms}$ when molecules that pass by the ODT can be cooled by sub-Doppler light into the trap. Since the ODT is a conservative trap, the enhanced loading suggests that cooling is occurring at least in some region of the trap. Another indication of sub-Doppler cooling of trapped molecules is that the measured temperature of $60(5)\,\mu$K is significantly lower than that expected from direct capture ($100\,\mu$K). To directly demonstrate that cooling is occurring in the presence of the trap light, we heat the CaF molecules to $\sim\!\!\,100$\,$\mu$K by applying a 40\,$\mu$s pulse of resonant light (detuned by $+3\,\text{MHz}$ to compensate for the Stark shift of the ODT beam) . To ensure that the heated molecules are trapped, we measure the molecule number after waiting $50\,\text{ms}$. After the resonant heating pulse, half of the trapped molecules remain. We then apply a pulse of sub-Doppler cooling light and observe that the molecules are re-cooled to $\sim\!\!\,60$\,$\mu$K with a $1/e$ time of $\sim\!\!\,300\,\mu$s (Fig.~\ref{fig:cooling}). This verifies that sub-Doppler cooling works for molecules trapped in the ODT. This is an important initial demonstration, as successful laser cooling of trapped molecules can pave the way towards highly efficient means to create dense ultracold samples. At the currently achieved densities, laser cooling is nearly lossless and, unlike evaporative cooling, is independent of collisional properties. Since entropy in the system is continuously and quickly removed, higher phase space densities can be reached with minimal loss. For example, by compressing the sample in the presence of cooling, the density could be increased with no rise in temperature.

In conclusion, we have demonstrated optical trapping of CaF ($X\,^2\Sigma$) molecules and laser cooling of the trapped samples to sub-Doppler temperatures. These results open up many possibilities. One immediately accessible application is the study of collisions of ground state molecules with atoms. One could explore the possibility of spin control of chemical reactions, e.g. between CaF and Li~\cite{lim15}, or proposed techniques such as collisional ``shielding"~\cite{bohn16}. Detailed characterization of inelastic loss channels and chemical reactions~\cite{Kosicki2017} can be carried out, which can improve our understanding of atom-molecule collisions, crucial for sympathetic cooling. With currently achieved trap lifetimes and further improvements in densities, the study of molecule-molecule collisions is within reach. Trapping of molecules in an ODT, as demonstrated in this work, can also produce sufficiently dense samples from which an optical tweezer can be loaded~\cite{schlosser2002tweezerblockade,yavuz2006tweezerrabi}. Laser cooling and photon cycling of optically trapped molecules is important for such experiments, as it enables optical readout of trapped single molecules, where many photons must be scattered with minimal heating. Extensions to arrays of optical tweezers of molecules would provide a pristine environment in which quantum simulation and quantum computation can be performed~\cite{lukin16array,barredo16array}. CaF is an ideal candidate for such explorations since it has a large electric dipole moment of 3\,Debye as well as a non-zero magnetic moment due to its unpaired electron spin. 
Optical trapping of laser-coolable molecules in tightly confining traps also opens up the possibility of Raman sideband cooling or electromagnetically-induced-transparency (EIT) cooling, both highly efficient cooling methods, which can circumvent the need for large initial numbers in evaporative cooling schemes~\cite{Vuletic1998, Roos2000, Kaufman2012, Thompson2013}.

\section{\label{summary}Acknowledgments}
This work was supported by NSF and ARO. BLA acknowledges support from NSF GRFP. LWC acknowledges support from MPHQ. We thank the Greiner group for lending us a 1064\,nm fiber amplifier.


\bibliography{motbib} 
\end{document}